\documentclass[10pt,prb,aps,twocolumn,superscriptaddress,showpacs,amsfonts,amsmath,amssymb,showkeys,floatfix]{revtex4-1}

\usepackage{bm}
\usepackage{pstricks}
\usepackage{setspace}
\usepackage{fdsymbol}
\usepackage{amssymb}
\usepackage{graphicx}

\begin{document}

\title{Phase diagram for ensembles of random close packed Ising-like dipoles  as a function of texturation   
}

\date{\today}
\author{Juan J. Alonso}
\email[e-mail address: ] {jjalonso@uma.es}
\affiliation{F\'{\i}sica Aplicada I, Universidad de M\'alaga, 29071 M\'alaga, Spain}
\affiliation{Instituto Carlos I de F\'{\i}sica Te\'orica y Computacional,  Universidad de M\'alaga, 29071 M\' alaga, Spain}
\author{B. All\'es}
\email[E-mail address: ] {alles@pi.infn.it}
\affiliation{INFN--Sezione di Pisa, Largo Pontecorvo 3, 56127 Pisa, Italy}
\author{V. Russier}
\email[E-mail address: ] {russier@icmpe.cnrs.fr}
\affiliation{ICMPE, UMR 7182 CNRS and UPE 2-8 rue Henri Dunant 94320 Thiais, France.}

\date{\today}

\begin{abstract}
We study random close packed systems of magnetic spheres by Monte Carlo simulations in order
to estimate their phase diagram. The uniaxial anisotropy of the spheres makes each of them behave
as a single Ising dipole along a fixed easy axis.  We explore the phase diagram 
in terms of  the temperature and the degree of alignment (or {\it texturation}) among the easy axes of all spheres. This degree of alignment ranges from the textured
 case (all easy axes pointing along a common direction)  to the non-textured case (randomly distributed easy axes). In the former case we find long-range ferromagnetic order at low temperature but,
as the degree of alignment is diminished below a certain threshold, the ferromagnetic phase gives way to a spin-glass phase. This spin-glass phase is similar to the one previously
found in other dipolar systems with strong frozen disorder. The transition between ferromagnetism and spin-glass passes through a narrow intermediate phase  with quasi-long-range ferromagnetic order.
\end{abstract} 
\maketitle

\section{INTRODUCTION}
\label{intro}

The study of ensembles of magnetic nanoparticles (NP) is an active field of research due to
its potential application in areas as disparate as biomedicine, data storage or nanofluids.\cite{np, bedanta}
Present technology allows to synthesize NPs with a wide variability of sizes and shapes, in addition to coating them with non-magnetic
layers. Moreover they can be produced  in nearly monodisperse ensembles so as to enjoy a good control on their spatial distribution.\cite{nano}
This know-how opens the possibility to realize densely packed ensembles of NPs that behave
as systems of interacting dipoles. It is the magnetic order of such structures that stirs a renewed interest in their use in technological applications.\cite{fiorani, sawako1}
 
NPs with diameters $d_p$ up to a few tens of nanometers have a single domain
(typical values are 15 nm for Fe, 35 nm for Co, 30 nm for maghemite $\gamma$-$Fe_{2}O_{3}$) 
 that behaves as a magnetic dipole.\cite{skomski}  Even when they are spherical, such NPs can have
anisotropies  that oblige the dipole to lie along a local easy axis and  to surmount
an anisotropy energy barrier $E_{a}$ whenever the magnetic moment is inverted, resulting in a
blocking temperature $T_{b} \simeq E_{a}/30 k_{B}$. \cite{bedanta, fiorani}
When the NPs  are closely packed,  their dipolar interaction energies $E_{dd}$ are not negligible but typically larger than $E_{a}/10$, leading to  $E_{dd}/ k_{B}T_{b} \gtrsim 3$. 
Consequently, low-temperature signatures of collective order induced by the dipolar interaction can be (and have indeed been) observed experimentally. \cite{toro1} 
This is to be compared with the super-paramagnetism observed in very diluted systems for which $E_{dd}/k_{B}T_{b} \ll 1$.\cite{bedanta, superpara} 

Dilute dispersions of NPs gather into highly ordered 3D super-crystals on account of their ability to self-assemble after the evaporation of the solvent.\cite{sc1,sc2}
Such crystals exhibit dipolar super-ferromagnetism in  FCC, BCC of I-tetragonal lattices. This behavior was predicted  to exist in such lattices by Luttinger and Tisza.\cite{lutti}

Less ordered (non-crystalline) dense packings may be obtained by pressing powders to obtain a granular solid,\cite{powder} or in concentrated colloidal suspensions 
by freezing the carrier fluid.\cite{ferrofluids} The frozen disorder on the positions of the NPs and on the orientation of the anisotropy axes in those systems
may induce frustration resulting in super spin-glass (SG) behavior.\cite{morup,russier} This behavior, originated by
dipolar interactions, has been observed experimentally in random close packed (RCP) samples of dipolar spheres\cite{toro1}  with volume fractions $\phi$ about 64\%.\cite{torquato} 
An equilibrium SG phase  for non--textured RCP ensembles of dipolar spheres has recently been found by numerical simulations.\cite{jpcm17} 

Nevertheless, the role of positional and orientational disorder in non-crystalline ensembles is far 
from being completely understood. Numerical simulations have shown that  frozen amorphous densely packed systems with volume
fractions as high as $\phi=0.42$ order ferromagnetically provided they are textured.\cite{ayton1, ayton2}
 This texturation shows up in colloidal suspensions by freezing the solution in the presence of large magnetic fields $h$.\cite{sawako2}
Even when $h=0$, ensembles of dipolar spheres moving in a non-frozen fluid with volume fractions as low as $42\%$ tend spontaneously
to become textured by aligning their axes, exhibiting nematic order (i.e. with no positional long range order). \cite{weis, weis2}

The picture that emerges is that the ordering of  dense non-crystalline systems may change
from  ferromagnetic (FM) to SG as the anisotropy-axes alignment dwindles from textured (i.e., parallel axes dipoles or PAD) to non-textured (random oriented axes dipoles or RAD).

The purpose of the present work is to depict the phase diagram of non-crystalline
dense packings of Ising dipoles with different degrees of texturation by employing Monte Carlo (MC) simulations (see Fig. \ref{phases}). 
In this effort, special attention will be paid to (i) examine whether a 
SG phase exists comparable to the one previously found for very diluted as well as RAD systems of Ising dipoles,
and (ii) explore the transition between FM and SG in order to look for possible intermediate phases.
We will pursue this investigation on ensembles of Ising dipoles placed at the center of RCP spheres that 
occupy a 64\% fraction of the entire volume.  Given that here we do not focus on time-dependent properties, we concede to the 
Ising dipoles (i.e. dipoles with large anisotropy energies) all the necessary time to flip up and down along their easy axes and reach equilibrium,
which is tantamount to say that we choose $T_{b}=0$. Such a model may be relevant for experimental situations  in which one expects $E_{a} \sim 10 E_{dd}$.\cite{toro1}
In order to investigate the effect of the easy axes alignment we will
introduce a parameter $\sigma$ that interpolates from the textured to the completely random axes cases.
The nature of the low temperature phases are investigated by measuring the spontaneous magnetization,
the SG overlap parameter, and the associated fluctuations and probability distributions.

The paper is organized  as follows. In Sec.~\ref{mm} we carefully define the model, give the details of the MC algorithm, 
and introduce the observables that will be measured. The results are presented in  Sec.~\ref{results} and some concluding remarks in Sec.~\ref{conclusion}.

\section{MODEL, METHOD, AND OBSERVABLES}
\label{mm}
\subsection{Model}
\label{models}
We study RCP systems of $N$ identical NPs that behave as single magnetic Ising dipoles. 
The NPs are labelled with $i=1,\dots, N$. We will regard each NP as a sphere of diameter $d$
carrying a permanent pointlike magnetic moment  $ \vec{\mu}_{i}=\mu s_{i} \widehat{a}_{i} $ at its center, where  the unit vector
 $\widehat{a}_i$ is the local easy-axis and $s_i=\pm1$.

The Hamiltonian governing the interaction is
\begin{equation}
{\cal H}= \sum_{ <i,j>}  \varepsilon_d\left( \frac {d}{r_{ij}} \right) ^{3}
\Big( \widehat{a}_i \cdot \widehat{a}_j-\frac {3(\widehat{a}_i\cdot \vec{r}_{ij})( \widehat{a}_j\cdot \vec{r}_{ij})} {r_{ij}^2}\Big) s_i s_j\;,
\end{equation}
where $\varepsilon_d =\mu_{0}\mu^2/(4 \pi d^{3})$ is an energy and $\mu_0$ the magnetic permeability in vacuum.
$\vec{r}_{ij}$ is the vector position of dipole $j$ viewed from dipole $i$, and $r_{ij}=\Vert\vec{r}_{ij}\Vert$.
The summation runs over all pairs of dipoles $i$ and $j$, with $i\not= j$.
The particles' positions as well as their easy axes $\widehat{a}_{i}$ remain fixed during the simulations.

The spheres are placed in frozen RCP configurations in a cube of edge $L$ assuming periodic boundary conditions. 
As in previous work,\cite{jpcm17} these configurations are obtained by using the Lubachevsky-Stillinger  algorithm,\cite{ls, donev}
in which the spheres, that are initially very small, are allowed to move and collide while growing in size at a sufficiently high rate until the sample gets eventually stuck in a non-crystalline state 
with volume fraction $\phi=0.64$.\cite{torquato, donev} We shall specify the size of the system by the
number $N$ of spheres inside it, or, equivalently by the lateral size of the cube they fill to capacity,
\begin{equation}
  L= \left(\frac{N \pi}{6\phi}\right)^{1/3}d\;.
  \label{equationdetomeu}
\end{equation}
where $d$ is the final diameter attained by the spheres after they ended growing.

To investigate the effect of texturation, we consider  that the alignment of the vectors
$\widehat{a}_{i}$ with the direction $\widehat{z}$ follows a Gaussian-like distribution
\begin{equation}
p(\theta_i) \propto \{ e^{-\theta_i^2/2{\sigma}^2   } + e^{-(\theta_i-\pi)^2/2{\sigma}^2}    \} \sin\theta_i, 
 \label{distri} 
\end{equation}
where $\theta_i$ is the polar angle of the $i$-th dipole while each azimuthal angle is chosen at random. The variance $\sigma$ controls the degree of texturation, intended as the amount of
alignment of the easy axes along the Cartesian axis $\widehat{z}$.  $\sigma$ ranges from  $\sigma=0$ for textured systems (PAD) 
to $\sigma=\infty$ for non-textured samples with  axes completely oriented at random (RAD).

We let each Ising dipole flip up and down along its easy axis $\widehat{a}_{i}$,
assuming that the dipoles are able to overcome the local anisotropy barriers. In what follows, distances and temperatures will be given in units 
of $d$ and $\varepsilon_d/k_{B}$ respectively,  where $k_{B}$ is the Boltzmann's constant.

\subsection{Samples}

We define a {\it sample} $\cal J$ as a given, arbitrary realization of disorder which, for the systems under study,
comes from two sources: from the randomness of the positions of the spheres and from the degree of texturation or of alignment of their easy axes $\widehat{a}_i$.
This disorder does not participate in the dynamics but remains frozen during MC simulations. Only the signs $s_i$ evolve during a simulation.

\begin{figure}[!t]
\begin{center}
\includegraphics*[width=84mm]{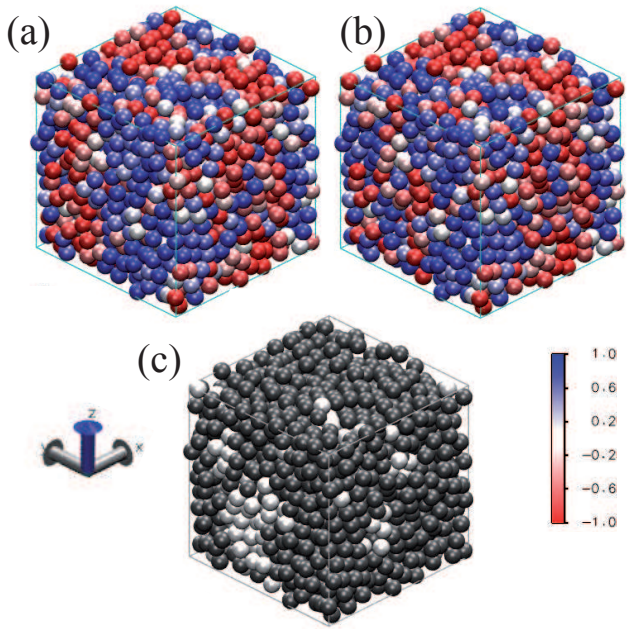}
\caption{(Color online)
(a) and (b) show two statistically independent configurations of a sample with
$1728$ magnetic nanospheres with  $\sigma=0.6$ at the  temperature $T=0.55$. 
The position of the spheres and the  orientation of their local easy axes are both frozen. 
The color of each sphere $i$ stands for the value of the $z$ component of
the  magnetic moment $ \vec{\mu}_{i}/\mu = s_{i} \widehat{ a}_{i}$, where 
$\widehat{a}_i$ is the local easy-axis and $s_i=\pm1$. Picture (c) represents the overlap
between the configurations (a) and (b). Black (white) color of spheres in (c) means
$s^{(a)}_{i}s^{(b)}_{i}=+1$ ($-1$).
}
\label{configsBR}
\end{center}
\end{figure}

As a consequence of the above definitions, we shall call {\it configuration} any set of $N$ signs $\{s_i\}_{i=1,\dots,N}$.
In Figs.~\ref{configsBR}(a,b) two statistically independent configurations obtained from a given sample by MC simulation are depicted.
Dark blue (red) colored spheres in the figures stand for dipoles pointing up (down) 
along axes $\widehat{a}_i$  nearly parallel to $\widehat{z}$, while light greyish spheres stand for those whose axes deviate significantly  from $\widehat{z}$.

Results susceptible to be compared with empirical data require an average over $N_{s}$ independent samples. The need of this average is crucial
at large $\sigma$ due to the sizeable sample-to-sample fluctuations that appear in this regime, where SG order is expected. Moreover, because of the lack of self-averaging  
associated with SG order,  we have not made $N_s$ smaller with increasing $N$. However, for large systems (the largest ones contain $N=1728$ dipoles)
we could employ no more than $3000$ samples because of computer time limitations. 
The number of samples $N_{s}$ is listed in Table~I  for the values of $N$ and $\sigma$ explored in the simulations.

\subsection{Method}

Since by decreasing the degree of texturation, the system could end up in a SG phase, we have performed parallel simulations with the tempered Monte Carlo (TMC) algorithm as this algorithm has proved to
be satisfactorily efficient in beating slowing down.\cite{tempered} Indeed, the TMC method allow replicas to overcome energy barriers within which the system could sink and remain confined
at low temperatures. These potential wells are minima of the rough energy landscapes that characterize glassy phases. Concretely, for each sample ${\cal J}$, we run in  parallel $n+1$ identical
replicas at  temperatures  $T= T_{\rm min}+k \Delta$ where $k=0,1,2,...,n$. We have found useful to choose the highest temperature, $T_{\rm max}=T_{\rm min}+n \Delta$, larger than
twice the transition temperature from the paramagnetic (PM) phase to the ordered one. The TMC algorithm involves two steps. In the first one, 10 Metropolis sweeps\cite{mc} are
applied separately  to all $n+1$ replicas, in order to make them evolve independently from each other. Dipolar fields are updated whenever a sign $s_j$ flip is accepted.
After that step, we give to any pair of replicas evolving at neighboring
temperatures $(T, T\pm\Delta)$ a chance to be exchanged, according to tempering rules that satisfy detailed balance.\cite{tempered} We choose $\Delta$ such that at least $30 \%$
of all attempted exchanges are accepted. Due to limitations in computer time we simulate systems containing up to $N=12^{3}=1728$ dipoles and choose $T_{\rm min}$ larger than half the transition temperature.

\begin{table}[!t]
\begin{tabular}{p{1.3cm} p{1.2cm } p{1.2cm } p{1.2cm } p{1.2cm }}
\hline
\multicolumn{5}{c}
{$\sigma=0$ ~~~($T_{\rm max}=4.5$,~~$T_{\rm min}=1.55$)}\\
 \hline
$N$ & $216$ & $512$ & $1000$  &1728\\
$N_s$ & $2100$ & $500$ & $500$  &500\\
\hline
\multicolumn{5}{c}
{$\sigma=0.1$ ~~~ ($T_{\rm max}=4.5$,~~$T_{\rm min}=1.55$)}\\
 \hline
$N$ & $216$ & $512$ & $1000$   &-\\
$N_s$ & $1000$ & $500$ & $500$   &-\\
\hline
\multicolumn{5}{c}
{$\sigma=0.2$ ~~~ ($T_{\rm max}=4$,~~$T_{\rm min}=1.05$)}\\
 \hline
$N$ & $216$ & $512$ & $1000$   &-\\
$N_s$ & $1000$ & $500$ & $500$   &-\\
\hline
\multicolumn{5}{c}
{$\sigma=0.3$ ~~~ ($T_{\rm max}=4$,~~$T_{\rm min}=1.05$)}\\
 \hline
$N$ & $216$ & $512$ & $1000$  &1728\\
$N_s$ & $1000$ & $2900$ & $2100$  &2000\\
\hline
\multicolumn{5}{c}
{$\sigma=0.4$ ~~~ ($T_{\rm max}=3.5$,~~$T_{\rm min}=0.55$)}\\
 \hline
$N$ & $216$ & $512$ & $1000$ &- \\
$N_s$ & $2000$ & $2000$ & $2000$   &-\\
\hline
\multicolumn{5}{c}
{$\sigma=0.45$ ~~~ ($T_{\rm max}=3.5$,~~$T_{\rm min}=0.55$)}\\
 \hline
 $N$ & $216$ & $512$ & $1000$ &$1728$ \\
$N_s$ & $10000$ & $2000$ & $2000$ & $2500$\\
\hline
\multicolumn{5}{c}
{$\sigma=0.50$ ~~~ ($T_{\rm max}=3.5$,~~$T_{\rm min}=0.55$)}\\
 \hline
$N$ & $216$ & $512$ & $1000$ &1728\\
$N_s$ & $10000$ & $8400$ & $6000$ &2000\\
\hline
\multicolumn{5}{c}
{$\sigma=0.53$ ~~~ ($T_{\rm max}=3.5$,~~$T_{\rm min}=0.55$)}\\
 \hline
$N$ & $216$ & $512$ & $1000$ & 1728\\
$N_s$ & $9800$ & $9600$ & $6500$ &2000\\
\hline
\multicolumn{5}{c}
{$\sigma=0.55$ ~~~ ($T_{\rm max}=3.5$,~~$T_{\rm min}=0.55$)}\\
 \hline
$N$ & $216$ & $512$ & $1000$ & 1728\\
$N_s$ & $10700$ & $8000$ & $4000$ & 2000\\
\hline
\multicolumn{5}{c}
{$\sigma=0.57$ ~~~ ($T_{\rm max}=3.5$,~~$T_{\rm min}=0.55$)}\\
 \hline
$N$ & $216$ & $512$ & $1000$ &1728\\
$N_s$ & $11600$ & $10300$ & $5000$ &3000\\
\hline
\multicolumn{5}{c}
{$\sigma=0.60$ ~~~ ($T_{\rm max}=3.5$,~~$T_{\rm min}=0.55$)}\\
 \hline
$N$ & $216$ & $512$ & $1000$ &1728\\
$N_s$ & $11000$ & $8000$ & $8400$ &8200\\
\hline
\multicolumn{5}{c}
{$\sigma=0.70, 0.80$ ~~~ ($T_{\rm max}=3.5$,~~$T_{\rm min}=0.55$)}\\
 \hline
$N$ & $216$ & $512$ & $1000$ &-\\
$N_s$ & $10000$ & $8000$ & $4800$ &-\\
\hline
\end{tabular}
\caption{The values taken by the parameters utilized in the TMC simulations. $\sigma$ is the degree of texturation,
 $N$ the number of dipoles, $N_{s}$ the number of samples with different realizations of disorder, and $T_{\rm max}$ and $T_{\rm min}$ the highest and lowest and temperatures respectively. $\Delta=0.05$ is the temperature step in all simulations. The number of MC sweeps for equilibration is $t_{0}=10^{6}$ in all cases. Measurements are taken during the MC sweeps comprised in the interval $[t_{0}, 2t_{0}]$.
 }
\label{table1}
\end{table}

We have imposed periodic boundary conditions in the simulations. That means that each dipole $i$ is allowed to interact with all dipoles within an $L\times L\times L$ box centered at $i$, see (\ref{equationdetomeu}).
Due to the long-range nature of the dipolar-dipolar interaction, we need to take into account contributions  from beyond this box by using Ewald's sums.\cite{ewald} Details on the use of Ewald's
sums for dipolar systems are given in Ref.\cite{holm}. In these sums, the use of neutralizing Gaussian distributions with standard deviation $\alpha/2$  allows
to split the computation of the dipolar fields into two rapidly convergent sums: a first sum in real space with a cutoff $r_{c}=L/2$, and a second sum in reciprocal space with a cutoff $k_{c}$. 
We have used $k_{c}=10$,  and  $\alpha=7.9/L$ as a good compromise between accuracy and computational speed.\cite{holm} More importantly, given that textured systems in our model
are expected to exhibit spontaneous magnetization at low temperatures, we have chosen the so-called conducting external conditions using surrounding permeability $\mu^\prime=\infty$,
in order to eliminate shape dependent depolarizing effects.\cite{weis,allen} 

The thermal equilibration times $t_{0}$ are assessed by the same procedure of Ref.\cite{jpcm17}.  The overlap $q(t)$ of configurations created from two replicas of the same
sample $\cal J$ are obtained by evolving the replicas independently after having started from random configurations. Then $t_0$ is the average over samples of
the value of $t$ at which $q(t)$ attains a plateau $q_{0}$ for each sample. In order to test the value thus obtained for $t_{0}$, we observed that a second overlap
$\widetilde q(t_{0},t_{0}+t)$ calculated for pairs of configurations of a single replica taken at times $t_0$ and $t_0+t$ remains stuck to $q_{0}$ as $t$ increases.\cite{PADdilu2}
It is found that the less textured the system is, the longer the equilibration time appears. This is due to the large roughness of the free-energy landscapes for non--textured systems.
For these hard-to-equilibrate systems, the overlap distributions $p_{\cal J}(q)$ exhibit numerous spikes
associated with the existence of several pure states.\cite{aspelmeier} In the simulations we have examined the $\pm q$ symmetry  
of the overlap distributions $p_{\cal J}(q)$ as an additional indication that all samples are well thermalized.\cite{jpcm17}

 A double average, the thermal one for each sample $\cal J$ and the above-mentioned average over the $N_{s}$ samples,  is needed to achieve physical results.
The first average is taken within the time interval $[t_0,2 t_0]$.  Given an observable $u$, the result of both averages will be symbolized by
$\langle u\rangle$. For simplicity, $\langle |u|^{p}\rangle$ will often be denoted by $u_{p}$.
The values of all the simulation parameters are listed in Table~I.

\subsection{Observables}
\label{meas}

The observables that have been measured in the course of the work are the following:

\begin{itemize}

\item[(i)] the specific heat $c$ from the fluctuations of the energy $e\equiv\langle {\cal H}\rangle/N$;

\item[(ii)] the $m_z$ component of the magnetization vector
\begin{equation} 
{\vec{m}} \equiv \frac{1}{N} \sum_i \widehat{a}_i s_i \;,
\label{m}
\end{equation}
as a way to characterize the FM behavior. Note that for a given sample, ${\vec{m}}$ 
does not rotate during the MC simulation. Rather, it aligns along the nematic director\cite{weis,allen}  $\widehat{\lambda}_{\cal J}$ that, for the model under study, is the eigenvector corresponding to the largest eigenvalue of the tensor
${\pmb{ \mathbb Q}}_{\cal J} \equiv  \frac{1}{2N} \sum_{i} (3 \widehat{a}_i \otimes \widehat{a}_i  -     \pmb{\mathbb{I}} )$.
Since ${ \mathbb Q}_{\cal J}$ is constant in time, $\widehat{\lambda}_{\cal J}$ remain frozen during the simulation.

We find that, for the values of $\sigma$ considered here, $\widehat{\lambda}_{\cal J}$ 
practically coincides with $\widehat{z}$. Then, it makes sense using $m_{z}$  as the FM order parameter instead of $\Vert\vec{m}\Vert$. 
In fact, we have also computed  $\Vert{\vec{m}}\Vert$ and their related quantities and found that they provide the same qualitative results that $m_{z}$.

\item[(iii)] The moments $m_{p}=\langle |m_{z}|^{p}\rangle$ for $p=1,2,4$,  that prove useful to calculate the magnetic susceptibility
\begin{equation}
\chi_{m}\equiv {N\over{k_{B}T}}(m_{2}-m_{1}^{2}),
\label{suscZ}
\end{equation}
and the dimensionless Binder cumulant 
\begin{equation}
B_{m}\equiv{1\over 2} (3-{m_{4}  \over m_{2}^{2} }).
\label{Bm}
\end{equation}

\item[(iv)]  As an useful tool to look for SG behavior, we calculate the overlap parameter,\cite{ea}
\begin{equation} 
q\equiv \frac{1}{N} \sum_i  s^{(1)}_i s^{(2)}_i\;,
\label{q}
\end{equation}
given a sample ${\cal J}$. $s^{(1)}_j$ and $s^{(2)}_j$ in this expression are the signs at site $j$ of two replicas of the given sample, denoted $(1)$ and $(2)$,
that evolve independently in time at the same temperature. Similarly as it has been done for $m_{z}$, we also measure $q_{p}\equiv \langle |q|^{p}\rangle$ for integer $p$,
and the corresponding Binder parameter $B_{q}\equiv{1\over 2} (3-{q_{4}  \over q_{2}^{2} })$.

\item[(v)] Finally, for each sample $\cal J$ we compute the probability distributions  $p_{\cal J}(m)$ and $p_{\cal J}(q)$, 
  as well as their average over samples, which will be denoted by $p(m)$ and $p(q)$.

\end{itemize}

Errors for all quantities are  obtained from the mean squared deviations of the sample-to-sample fluctuations.

\section{RESULTS}
\label{results}

\subsection{The FM Phase}
\label{FMphase}

\begin{figure}[!t]
\begin{center}
\includegraphics*[width=75mm]{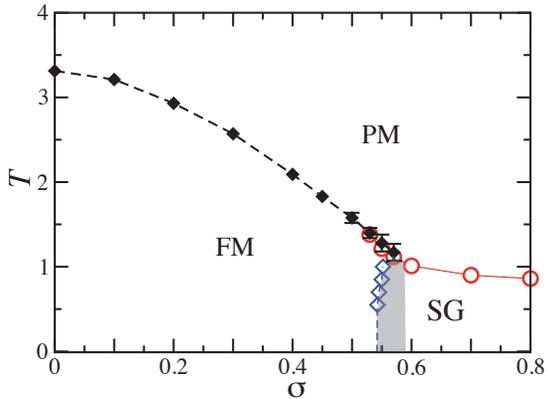}
\caption{(Color online) Phase diagram on the temperature-degree of texturation plane for the dipolar Ising model. Symbols $\blackdiamond$ indicate the PM-FM
  transition and have been obtained from data of $B_{m}$ vs $T$. Symbols $\medcircle$ stand for PM-FM and PM-SG transitions
  and were obtained from the $B_{q}$ vs $T$ plots. Symbols $\meddiamond$ represent the FM-SG transition and follow
  from the $B_{m}$ vs $\sigma$ plots. The error bars for the data marked with $\medcircle$ and $\meddiamond$ are smaller than the size of these symbols.
  FM quasi-long-range order cannot be discarded in the grey region.}
\label{phases}
\end{center}
\end{figure}

The main result of the paper is the phase diagram on the plane temperature-degree of texturation shown in Fig. \ref{phases}.  It displays regions with FM, PM and SG phases.
The FM order arises at low temperatures in the range $0 \le \sigma \lesssim 0.53$. A thermally driven second order transition takes place at 
the phase boundary between the PM and FM phases. Next we give the numerical evidence that supports this interpretation.

\begin{figure}[!t]
\begin{center}
\includegraphics*[width=84mm]{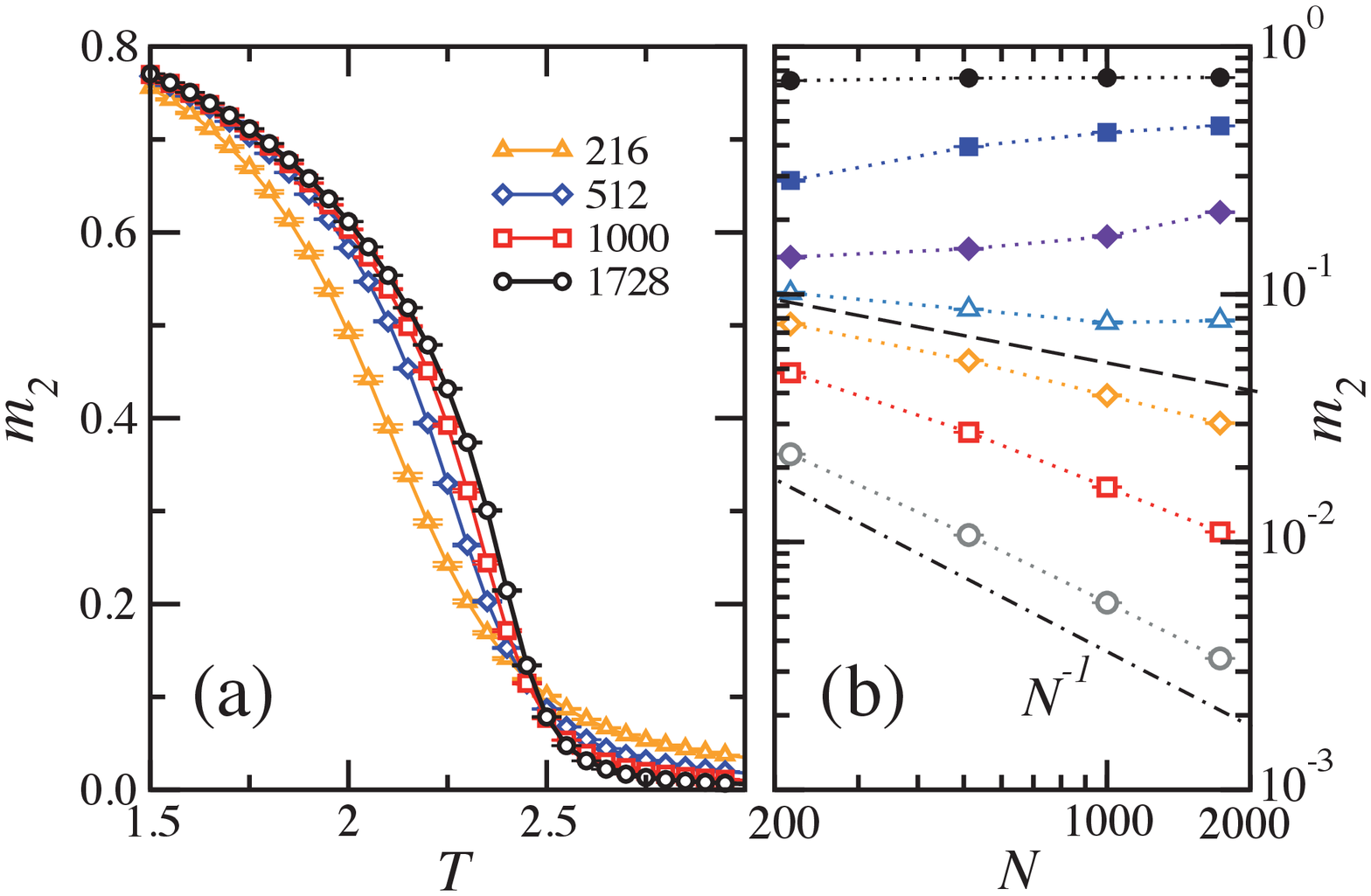}
\caption{ (Color online)  
(a)  Plots of the squared magnetization $m_{2}$  vs temperature $T$ for $\sigma=0.3$. 
$\smalltriangleup$,  $\smalldiamond$, $\square$ and $\smallcircle$  stand for $N=216, 512, 1000$ and $1728$ dipoles respectively. Lines are guides to the eye.
(b) Log-log plots of $m_2$  vs $N$ for different temperatures at $\sigma=0.3$. From top to bottom,
$\smallblackcircle$, $\smallblacksquare$, $\blackdiamond$, $\smalltriangleup$,  $\smalldiamond$, $\square$ and $\smallcircle$
stand for $T=1.6,~ 2.2,~ 2.4,~ 2.5, ~2.6, 2.8$ and $3.4$ respectively. 
Dotted lines are guides to the eye. The dashed line separates two regimes and stand for a $1/N^{0.35}$ decay.
The dot-dashed line shows the $N^{-1}$ decay expected for paramagnets in the thermodynamic limit.}
\label{magne30}
\end{center}
\end{figure}

\begin{figure}[!b]
\begin{center}
\includegraphics*[width=80mm]{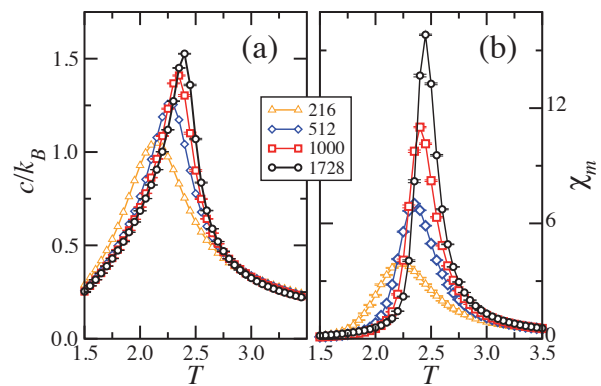}
\caption{
(Color online)
(a) Plots of the specific heat  versus  $T$ for $\sigma=0.3$. 
$\smalltriangleup$,  $\smalldiamond$, $\square$ and $\smallcircle$  stand for systems with $N=216, 512, 1000$ and $1728$ dipoles respectively. 
(b) Plots of the magnetic susceptibility $\chi_{m}$  vs $T$  for $\sigma=0.3$. Same symbols as in (a).  Lines in both panels are guides to the eye.
}
\label{heat-chi-30}
\end{center}
\end{figure}

FM phases are defined by the presence of a non-vanishing magnetization. In Fig. \ref{magne30}(a) we show the behavior
of the moment $m_2$ with the temperature for $\sigma=0.3$ in a number of system sizes.  We obtain similar 
results for the magnetization for all values of $\sigma$  below $0.53$. This is a first piece of evidence of the existence of the FM phase. 
Fig.~\ref{heat-chi-30}(a) shows plots of the specific heat $c$ vs $T$. The sharp variation of $c$ near $T=2.5$ suggests the presence of a singularity
as $N$ increases, as it is expected for a second order PM-FM phase transition. 
The same happens with the plots of the magnetic susceptibility $\chi_{m}$ vs $T$ shown in Fig.~\ref{heat-chi-30}(b).
The data are consistent with a logarithmic divergence of $c$, and with an approximate power-law divergence of $\chi_{m}$ with  $N^{p}$  (up to logarithmic corrections $\ln N$) where $p \sim 2/3$. 

Next we examine the dependence of $m_2$ on the number $N$ of dipoles. Fig. \ref{magne30}(b) shows
log-log plots of $m_2$ vs $N$  for  several temperatures. The data at $T$ below
$T_{c}= 2.55(5)$ reflect that $m_2$ does not vanish in the $N \to \infty$ limit.
On the contrary the plot of $m_2$ vs $N$ for $T > T_{c}$ shows  a faster than a power-law decay with a $T$-dependent exponent, and
consequently  the slope of the curves is steeper for increasing $T$
and approaches a $1/N$ trend, which is the expected trend in PM phases.
The dashed line in Fig.\ref{magne30}(b) separating the two regimes represents a  $1/N^{0.35}$ decay.
Although we are aware that these graphs do not allow a precise determination of $T_{c}$,
we have followed this criterion as a first rough approach for establishing the boundary of the FM phase.

\begin{figure}[!b]
\begin{center}
\includegraphics*[width=84mm]{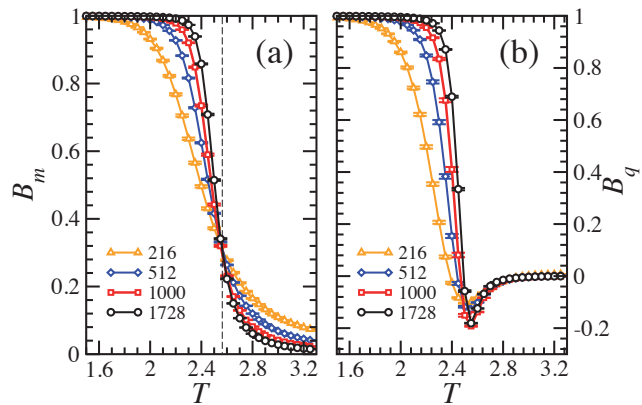}
\caption{(Color online)
  (a) Plots of the Binder cumulant $B_{m}$  vs  $T$ for $\sigma=0.3$. 
  $\smalltriangleup$,  $\smalldiamond$, $\square$ and $\smallcircle$  stand for systems with $N=216, 512, 1000$ and $1728$ dipoles respectively.
  The dashed vertical line indicates the Curie temperature, at which curves cross.
  (b) Plots of the Binder cumulant for the overlap parameter $B_{q}$  vs $T$. Same symbols as in (a). Solid lines in both panels are guides to the eye.
}
\label{binders30}
\end{center}
\end{figure}

\begin{figure}[!b]
\begin{center}
\includegraphics*[width=70mm]{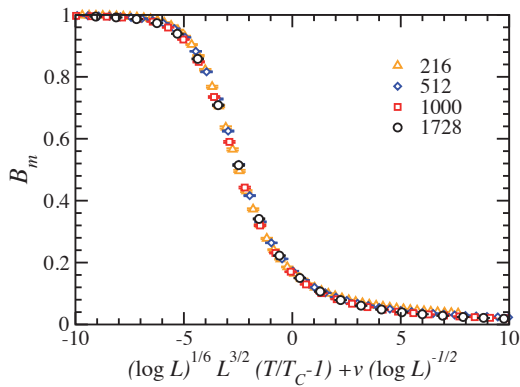}
\caption{ 
(Color online) Finite size scaling plots for $B_{m}$ vs
$L^{3/2}~(\log L)^{1/6}~(T/T_{c}-1)+v~(\log L)^{-1/2}$
for $\sigma=0.3$ using $T_{C}=2.57(2)$ and $v=-1.83(8)$.
$\smalltriangleup$,  $\smalldiamond$, $\square$ and $\smallcircle$  stand 
for systems with $N=216, 512, 1000$ and $1728$ respectively.
}
\label{fss30}
\end{center}
\end{figure}

The Binder parameter $B_m$ grants a more precise determination of the transition
temperature. It follows from its definition in (\ref{Bm}) that $B_{m}\rightarrow1$ as $N\rightarrow \infty$ in the FM phase.
On the other hand, from the law of large numbers it follows that, in the PM phase, with short-range FM order, $B_{m}\rightarrow0$ as $N$ increases.  
Finally, at a critical point, $B_{m}$ becomes size independent, as it must occur to every scale-free  observable (recall that $B_m$ is dimensionless).
The latter is also true in the case of a marginal phase with quasi-long-range magnetic order. 
Then, curves of $B_m$ vs $T$ for various values of $N$ should cross at $T_{c}$ if it is a second
order transition. Note however that when a marginal phase exists these curves should colapse rather than cross for all the critical region.\cite{balle}

The plots of $B_{m}$ vs  $T$ are shown in Fig.~\ref{binders30}(a) for different values of $N$ at $\sigma=0.3$.
It is apparent that all curves intersect at a precise temperature, allowing to extract the Curie temperature
$T_{c}(\sigma)$, and permitting to establish a clear-cut boundary between the PM and FM phases. 
The relatively modest system sizes that we have used (a limitation due to the long-range  nature of the dipolar interaction) does not allow the precise determination of the critical exponents. 

However, from finite size scaling relevant for dipolar Ising models we get acceptable data-collapse plots of
$B_{m}$ vs $L^{3/2}~\log^{1/6}L~(T/T_{c}-1)+v~(\log L)^{-1/2}$, that provide a more reliable determination of $T_{c}$,  (see Fig.~\ref{fss30}).
This finite size scaling behavior corresponds to the mean field one and agrees with the fact that the upper critical dimension of the dipolar Ising model be $d_u = 3$. \cite{aha,klopp}
For $\sigma=0.3$, we get $T_{c}=2.57(2)$.
Likewise, precise determinations of $T_c(\sigma)$ can be obtained for $\sigma \le 0.53$, the overall result being shown in Fig.~\ref{phases}.

For $\sigma = 0.55$ and $0.57$ the curves $B_{m}$ vs $T$ merge rather than cross at low temperatures,
giving a less precise determination of $T_{c}$. We will return to this point in subsection \ref{FMSG}.
Given that for our model  ${\vec{m}}$ does not rotate, $m_{z}$ 
and  the overlap $q$ are expected to give similar information in the FM phase. 
Thus, crossing  points in the plots of $B_{q}$ vs $T$  like the ones shown in Fig. \ref{binders30}(b), may in principle provide
an additional way for obtaining $T_{c}$. This is true for $\sigma \ge0.53$ for which clean crossing 
points are obtained. For smaller values of $\sigma$, see Fig.~\ref{binders30}(b), a
characteristic dip near the transition temperature makes it difficult to accurately locate the critical point.\cite{korean}

\subsection{The SG phase}
\label{SG}

This subsection is devoted to the study of small texturations, which quantitatively entails large values of $\sigma$.
As $\sigma$ grows, we observe large sample-to-sample fluctuations which obliges us to increase the number of samples up to roughly ten thousand (see Table~I) in order to
attain trustworthy averages. Also large relaxation times are observed, a typical feature of SG behavior.
 Indeed, we are going to report numerical data that evidence the absence of magnetic order and the existence of an equilibrium SG phase for systems with $\sigma \ge 0.6$.
With the aim of exploring this low-temperature ordered phase within a reasonable amount of computer time, we have performed the TMC simulations at temperatures
  no less than $T=0.55$ and system sizes no larger than $N=1728$, to the detriment of the accuracy.

\begin{figure}[!t]
\begin{center}
\includegraphics*[width=84mm]{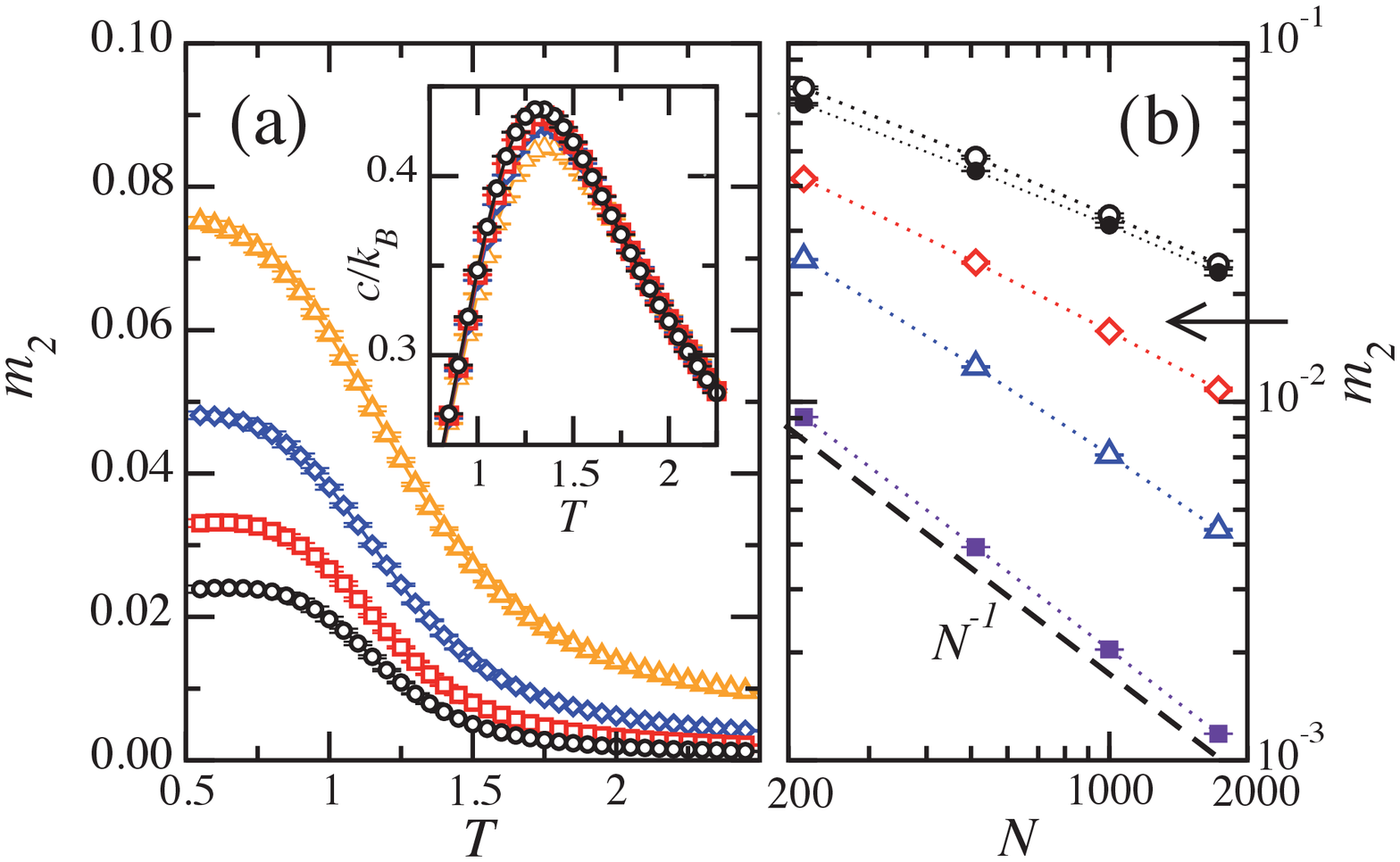}
\caption{(Color online) 
(a) Plots of the squared magnetization $m_{2}$ vs $T$ for $\sigma=0.6$. In the inset, the specific heat vs $T$.
$\smalltriangleup$,  $\smalldiamond$, $\square$ and $\smallcircle$  stand for $N=216, 512, 1000$, and $1728$ dipoles respectively. All lines in this panel are guides to the eye.
(b) Log-log plots of $m_2$ vs $N$ for $\sigma=0.6$. From top to bottom, $\smallcircle$, $\smallblackcircle$, $\smalldiamond$, $\smalltriangleup$, and $\smallblacksquare$ 
stand for temperatures $T=0.55,~ 0.85,~ 1.25, ~1.55$, and $2.55$ respectively. The arrow marks the onset of the PM phase.
Dotted lines are guides to the eye. The dashed line shows the $N^{-1}$ decay expected for a paramagnet in the thermodynamic limit.
}
\label{magne60}
\end{center}
\end{figure}

Plots of the moment $m_2$ vs $T$ are shown in Fig.~\ref{magne60}(a) at $\sigma=0.6$. $m_2$ decreases as $N$ increases at all temperatures.
In the inset of the figure, we show the plots of the specific heat $c/k_{B}$ vs $T$. They display a gentle variation and no signature of any possible
singularity is seen. Similar graphs follow if the study is repeated at larger values of $\sigma$. These are the first
pieces of evidence that point to the non-existence of FM order and of any PM-FM transition for $\sigma \ge 0.6$. 

In Fig.~\ref{magne60}(b) we show log-log plots of $m_2$ vs $N$. They exhibit a decay
faster than $1/N^{1/2}$ for all available temperatures. At low temperatures $T\lesssim 1$  the results are in principle
consistent with quasi-long-range magnetic order. We will further discuss this point in the next subsection.
For the PM phase (with short-range  magnetic order), we expect to observe  $m_2 \sim 1/N$ for 
large enough systems. For the available system sizes, we discern such a trend only for extremely large
temperatures, (see for example the data at $T=2.55$).

\begin{figure}[!b]
\begin{center}
\includegraphics*[width=80mm]{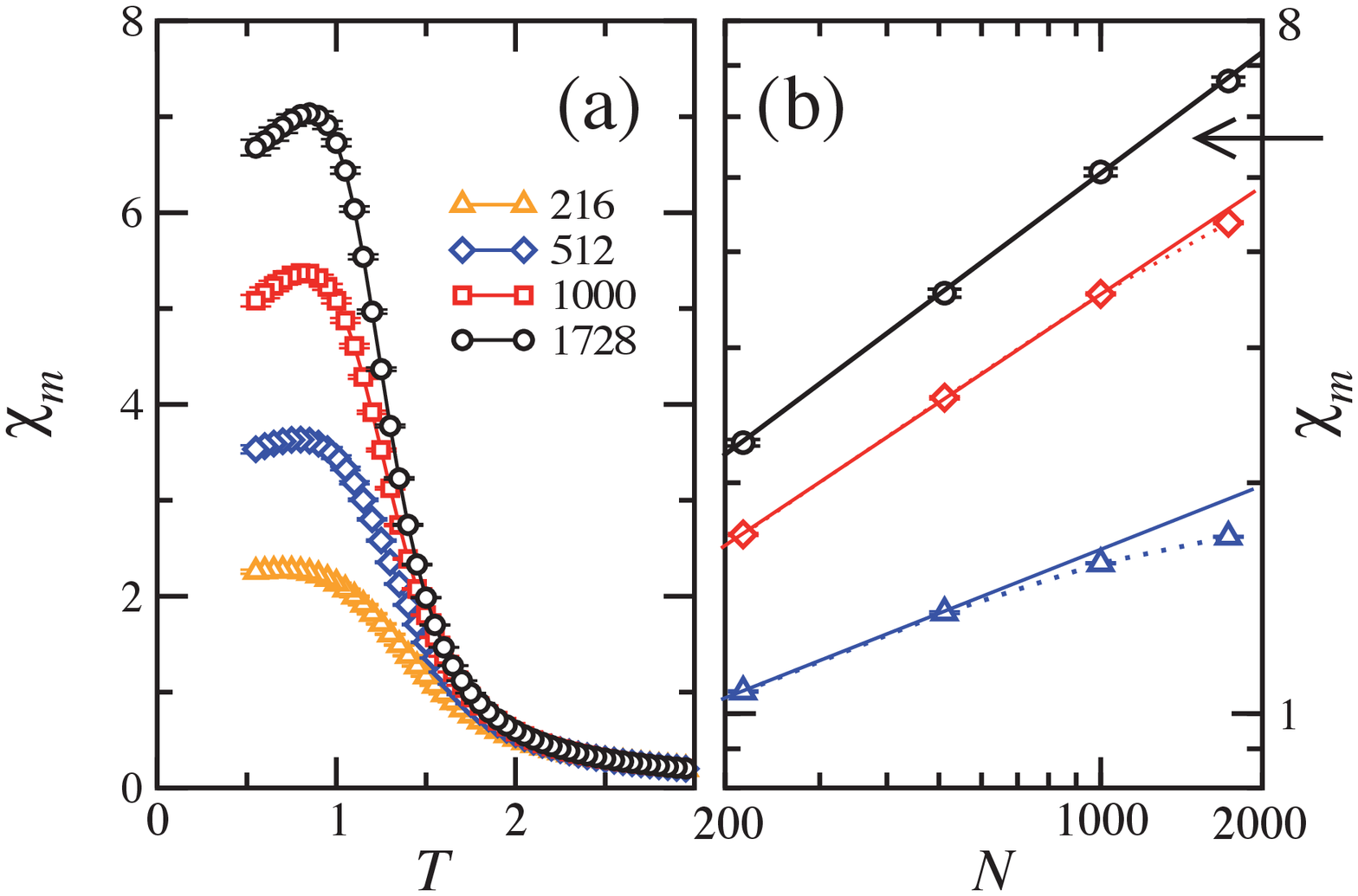}
\caption{ (Color online) 
(a) Plots of the magnetic susceptibility $\chi_{m}$  vs $T$ for $\sigma=0.6$. 
$\smalltriangleup$,  $\smalldiamond$, $\square$ and $\smallcircle$  stand for systems with $N=216, 512, 1000$ and $1728$ dipoles respectively.  Solid lines are guides to the eye.
(b) Log-log plots of $\chi_{m}$  vs  $N$ for $\sigma=0.6$. $\smallcircle$, $\smalldiamond$, and $\smalltriangleup$ stand for temperatures $T=0.55,~ 1.25$, and $1.55$ respectively.
As stressed by the dotted lines connecting the points, data ceases to grow linearly (the solid lines) at large temperatures. The arrow marks the onset of the PM phase.
}
\label{susce60}
\end{center}
\end{figure}

A definite signature of the presence of a SG phase is the divergence of the magnetic susceptibility at low temperatures.
The plots of $\chi_{m}$ vs $T$ for $\sigma=0.6$ showing an increase with $N$, see Fig.~\ref{susce60}(a), are consistent with that scenario.
Notice that this is in clear contrast with the behavior shown in  Fig.~\ref{heat-chi-30}(b) for $\sigma=0.3$. 
 Log-log plots of $\chi_{m}$ vs $N$ for low temperatures show a power-law increase $\chi_{m}\sim N^p$ with an exponent  $p$ that changes
slightly with $T$ but that is never greater than $p=0.55$ (see Fig.~\ref{susce60}(b)).
For $T \gtrsim1.25$,  the curves detach from an algebraic growth and bend downwards
indicating a non-diverging $\chi_{m}$ in the macroscopic limit, as expected for a PM phase.

\begin{figure}[!t]
\begin{center}
\includegraphics*[width=84mm]{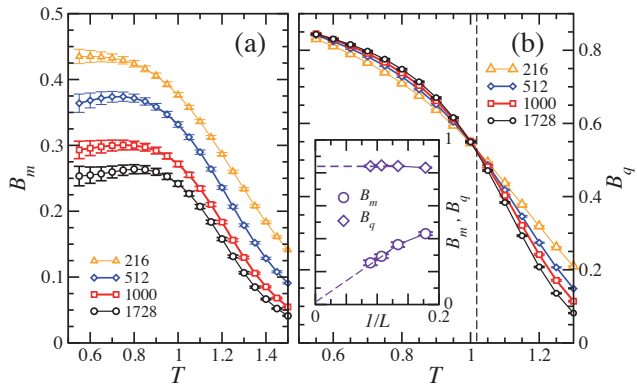}
\caption{(Color online) 
  (a) Plots of  $B_{m}$  vs  $T$ for $\sigma=0.6$. 
  $\smalltriangleup$,  $\smalldiamond$, $\square$ and $\smallcircle$  mean $N=216, 512, 1000$ and $1728$ dipoles respectively.  The solid lines are guides to the eye.
  (b) Plots of $B_{q}$  vs $T$. Same symbols as in (a). The curves cross at the SG transition temperature, marked in the figure with a vertical dashed line.
  The inset contains plots of $B_{m}$ and $B_{q}$ vs $1/L$ for the lowest temperature available, $T=0.55$. $\smallcircle$ ($\diamond$) stands for $B_{m}$ ($B_{q}$). The dashed lines in the inset
  are extrapolations. }
\label{binders60}
\end{center}
\end{figure}

The most convincing evidence for the absence of FM order at low temperatures for $\sigma=0.6$ is given in Fig.~\ref{binders60}(a). The $B_{m}$
vs $T$ plots show that $B_{m}$ diminishes as $N$ increases for all temperatures. As a consequence, curves for different system
sizes do not cross, in contrast with the behavior found in Fig.~\ref{binders30}(a). Recall that, in case of short-range 
FM order, $B_{m}$ should vanish in the thermodynamic limit. In the inset of Fig.~\ref{binders60}(b), we have represented $B_{m}$ vs $1/L$ for $T=0.55$,
showing that that is indeed the case. We obtain a similar trend for all $\sigma \ge 0.6$ and temperatures. 
This finding, consistent with short-range FM order, 
seems to be in contradiction with the effective power-law decay of $m_{2}$ with $N$ observed for low $T$ for the system sizes we have used (see Fig.~\ref{magne60}(b)).  Some clues could
be obtained by 
inspecting the two independent magnetic configurations displayed in  Fig.~\ref{configsBR}.
These are thermalized configurations at $\sigma=0.6$, $T=0.55$ in the largest system size considered in this work, $N=1728$. 
The sample appears to be broken into large magnetic domains whose frontiers appear to be frozen. The large
size of the domains explains the effective power-law decay found in the $m_{2}$ vs $N$ plots in Fig.~\ref{magne60}(b). 
In striking contrast, the overlap between the two configurations covers practically the whole system (see Fig.~\ref{configsBR}(c)), 
suggesting a diverging SG overlap correlation length.

\begin{figure}[!t]
\begin{center}
\includegraphics*[width=88mm]{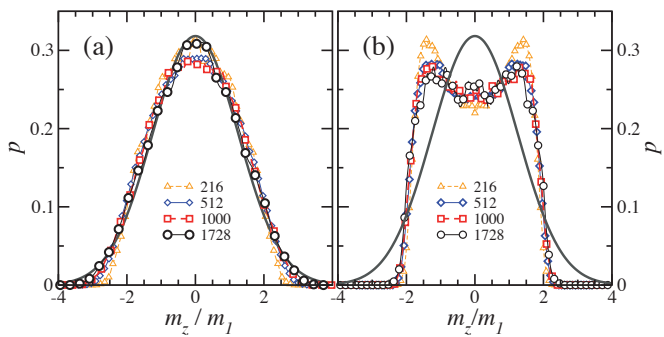}
\caption{(Color online) (a) Plots of the probability distribution $p(m_{z}/m_{1})$ for $\sigma=0.6$ and
  $T=0.55$. $\smalltriangleup$,  $\smalldiamond$, $\square$ and $\smallcircle$  stand for  $N=216, 512, 1000$ and $1728$ respectively.
  The thick solid line is the typical Gaussian distribution for paramagnets in the $N \to \infty$ limit.
  (b) Same as in (a) but for $\sigma=0.55$.  The thin lines connecting data in both panels are guides to the eye.
  }
\label{distribs}
\end{center}
\end{figure}

\begin{figure}[!b]
\begin{center}
\includegraphics*[width=80mm]{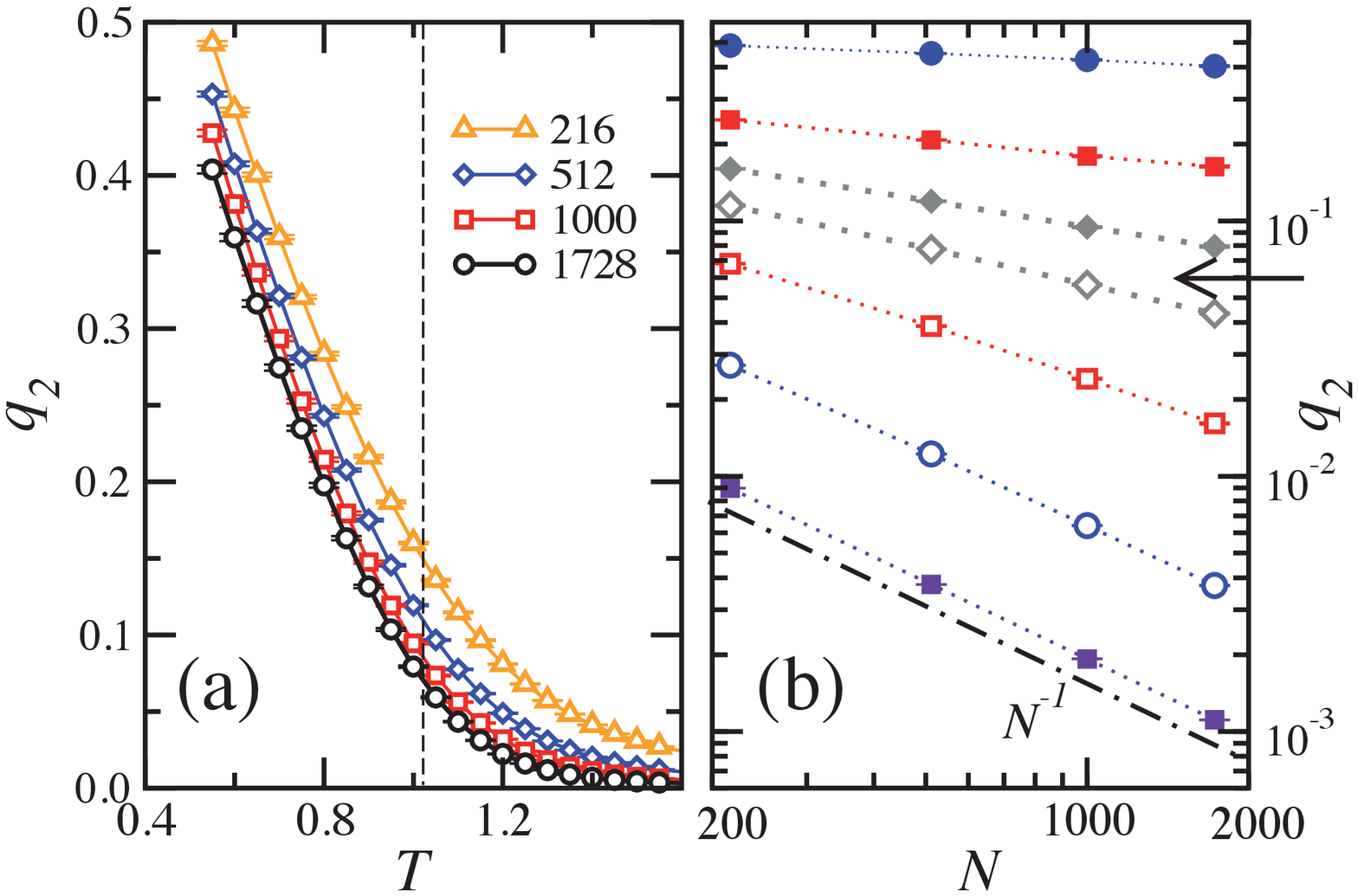}
\caption{(Color online) 
(a) Plots of the squared overlap parameter $q_{2}$  vs $T$ for $\sigma=0.6$. 
  $\smalltriangleup$,  $\smalldiamond$, $\square$ and $\smallcircle$  stand for  $N=216, 512, 1000$ and $1728$ respectively. The dashed vertical line indicates the SG transition temperature
 Solid lines are guides to the eye. (b) Log-log plots of $q_2$  vs the number of dipoles $N$ for $\sigma=0.6$. 
From top to bottom, $\smallblackcircle$, $\smallblacksquare$, $\smallblackdiamond$, $\smalldiamond$, $\square$, $\smallcircle$, and $\smalltriangleup$
stand for $T=0.55,~ 0.85,~ 1.0, ~1.05, ~1.25, ~1.55$, and $2.55$ respectively. The arrow marks the onset of the PM phase. 
Dotted lines are guides to the eye. The dot-dashed line shows the $N^{-1}$ decay expected for the PM phase.
}
\label{overlap60}
\end{center}
\end{figure}

Provided that the magnetic correlation length (i.e. the size of the magnetic domains) does not diverge, then
$m_{z}$ would be expected to be normally distributed, as follows from the law of large numbers. In Fig.~\ref{distribs}(a)
we represent the distribution $p(m_{r})$ where $m_{r}\equiv{m}_{z}/m_{1}$ averaged over all samples for $\sigma=0.6$ and 
the lowest temperature available, $T=0.55$. Clearly, $p(m_{r})$ tends to $(1/\pi) \exp (-m_r^2/\pi )$ as 
$N\rightarrow \infty$, in agreement with short-range magnetic order. We obtain qualitatively similar results for all $\sigma \ge 0.6$ and $T$,
a fact that leads us to discard the existence of a critical FM phase with quasi-long-range order at low temperature. For this to be the case, we should have seen
a non-Gaussian broad distribution $p(m_{r})$ that behaves as an scaling function that does not change with the system size.\cite{criti} It seems to be
the case, within errors, for a bit larger texturation ($\sigma=0.55$), as shown in Fig.~\ref{distribs}(b) for $T=0.55$. 
More details on this point will be discussed in the next subsection.

Finally, we report numerical evidence in favor of the positive existence of a SG phase for $\sigma \gtrsim 0.6$ by studying the overlap parameter $q_2$ and $B_{q}$.
Plots of $q_2$ vs $T$ are shown in Fig.~\ref{overlap60}(a) for $\sigma=0.6$.
It is worth comparing this figure with its counterpart for $m_{2}$, Fig.~\ref{magne60}(a), to appreciate the qualitative differences between the behavior of $q_2$ and $m_2$ at low temperature.
Note however that $q_2$ also decreases appreciably as $N$ increases for all temperatures.
This fact raises the question on whether or not $q_2$ vanishes as $L\rightarrow \infty$. To clarify this, we have prepared the log-log plots of $q_2$ vs $N$ shown in Fig.~\ref{overlap60}(b).
Data are consistent with $q_2\sim 1/N^{p}$ for low temperatures, and with a $T$--dependent exponent $p$.
The $N^{-1}$ trend, expected for PM phases, shows up only at large temperatures.  All of this suggests the presence of a phase with quasi-long-range SG order.
We draw additional evidence on this point from the behavior of $B_{q}$. Recall that in the thermodynamic limit 
$B_{q} \to 1$ in case of strong long-range order, vanishes in the PM phase, and tends to some
intermediate value at criticality. In Fig.~\ref{binders60}(b), plots of $B_{q}$ versus $T$ for $\sigma=0.6$ show
that curves of different system sizes cross at a precise temperature $T_{sg}$ that delimits the extend of the region with SG order. 
These crossings permit to obtain the points $T_{sg}(\sigma)$ of the PM-SG  transition line in Fig.~\ref{phases}. \cite{extraSG}
Note that $T_{sg}$ does not vary strongly with $\sigma$. The results agree well with the 
limiting value $T_{sg}= 0.8$ found in previous work for the RAD case ($\sigma=\infty$).\cite{jpcm17}
It is important to stress that the fact that the $B_{q}$ curves cross at $T_{sg}$
does not imply the existence of strong long-range order for $T \less T_{sg}$.\cite{PADdilu}
Indeed, plots of  $B_{q}$ vs $1/L$  for  $T \le T_{sg}(\sigma)$ show that $B_{q}$ stays below~1
(see the inset in Fig.~\ref{binders60}(b)). Then, the $B_{q}$ curves should collapse in the
$N \to \infty$ limit when $T \le T_{sg}(\sigma)$, which is consistent with the algebraic decay found for $q_{2}$. 

In summary, the data for $\sigma \ge 0.6$ point to the existence of a SG  phase 
delimited by $T_{sg}(\sigma)$ for which quasi-long-range SG order occurs, like in the 2D XY model.\cite{xy,xy2} A similar SG phase has been previously found  for other dipolar systems with strong frozen disorder,
namely for systems of parallel Ising dipoles  with strong dilution\cite{PADdilu, PADdilu2} as well as in 
dense arrays, both crystalline of not, of  non-textured systems of Ising dipoles with the axes oriented 
completely  at random.\cite {jpcm17,RADjulio} 
However, given the moderate range of system sizes considered here, our data cannot rule out completely the so-called
replica symmetry breaking  scenario in  which $q_{2}$ does not vanish in the $N \to \infty$ limit, but there are
long range SG order fluctuations which provoke $B_{q} < 1$.\cite{RSB, bookstein}

\begin{figure}[!b]
\begin{center}
\includegraphics*[width=80mm]{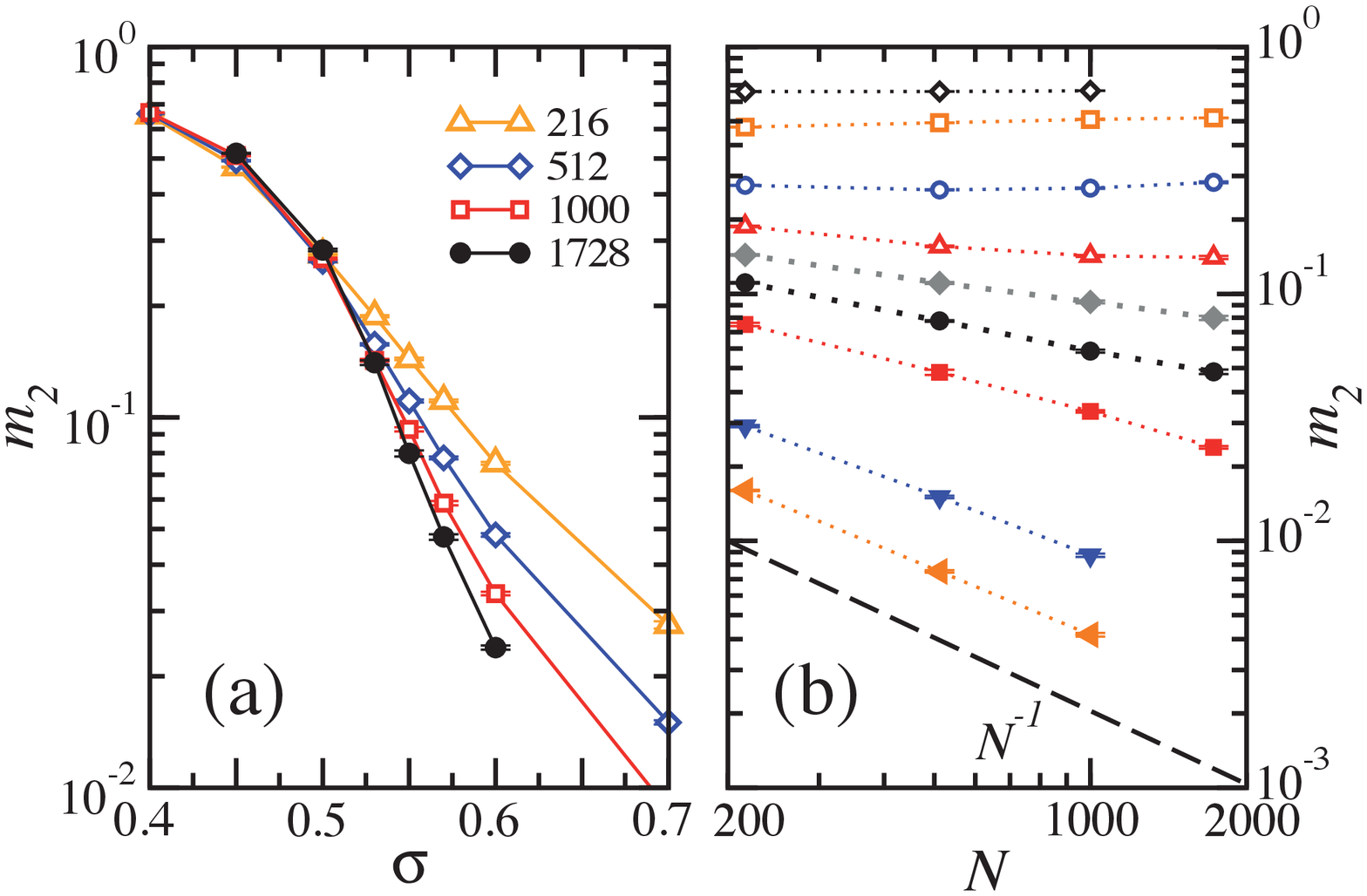}
\caption{ (Color online) 
(a) Semilog plots of the squared magnetization $m_{2}$  vs $\sigma$ for the lowest available temperature, $T=0.55$.  $\smalltriangleup$,  
$\smalldiamond$, $\square$ and $\smallcircle$ stand for  $N=216, 512, 1000$ and $1728$ respectively. Solid lines are guides to the eye.
(b) Log-log plots of $m_2$  vs the number of dipoles $N$ at $T=0.55$. 
From top to bottom, $\smalldiamond$, $\smallsquare$, $\smallcircle$, $\smalltriangleup$, $\smallblackdiamond$, $\smallblackcircle$,
$\smallblacksquare$,  $\smallblacktriangledown$, and $\smallblacktriangleleft$ 
stand for $\sigma=0.4,~ 0.45,~ 0.5, ~0.53, ~0.55, ~0.57,~0.6, ~0.7$, and $0.8$ respectively. 
Dotted lines are guides to the eye. The dashed line shows the $N^{-1}$ decay expected for the paramagnetic
phase.
  }
\label{m2-SIG}
\end{center}
\end{figure} 

\subsection{The FM-SG transition.}
\label{FMSG}

From the previous sections, we expect to find a transition within the narrow  region $0.53 \less \sigma \less 0.6$.
In order to identify it, we have carried out TMC simulations for several values of $\sigma$ in the interval $[0.45, 0.6]$
and a range of temperatures in the TMC between $T_{\rm max}=3.5$ and $T_{\rm min}=0.55$.
The highest temperature has been chosen well into the PM phase in order to refresh configurations and 
ensure equilibrium results for  $T_{\rm min}=0.55$  which is, in turn, a temperature 
well deep  into the low-temperature phase. This procedure facilitates the exploration of the FM boundary along several isothermal lines, allowing to investigate
whether there is an intermediate phase between this boundary and the SG phase determined in the
previous section.  In addition, the  slope of the FM boundary line may discern between a forward or a reentrant behavior.

The magnetization $m_{2}$ vs $\sigma$ in Fig.~\ref{m2-SIG}(a) for a low temperature $T=0.55$
shows that  $m_{2}$ decreases with $N$  for $\sigma > 0.5$.
Log-log plots of $m_{2}$ vs $N$ in Fig.~\ref{m2-SIG}(b) show that the $m_{2}$ curves deviate from an algebraic 
decay to bend upwards at $\sigma = 0.53$, indicating also a non-vanishing magnetization. 
In contrast, for  $\sigma = 0.55$ and $0.57$ we find a power-law
decay, giving some room for the existence of an intermediate region with quasi-long-range FM order. 
This decay is consistent with the behavior found for the $p(m_r)$ distributions of Fig.~\ref{distribs}(b)
for $\sigma=0.55$. All $p(m_r)$ curves tend to collapse into a non-Gaussian broad distribution for large $N$,
as expected when quasi-long-range order settles. We obtain the same qualitative
results for $\sigma=0.57$. Finally, curves for larger values of $\sigma$ tend to the  $N^{-1}$ 
decay characteristic of short-range FM order, as discussed in the previous section.

\begin{figure}[!b]
\begin{center}
\includegraphics*[width=80mm]{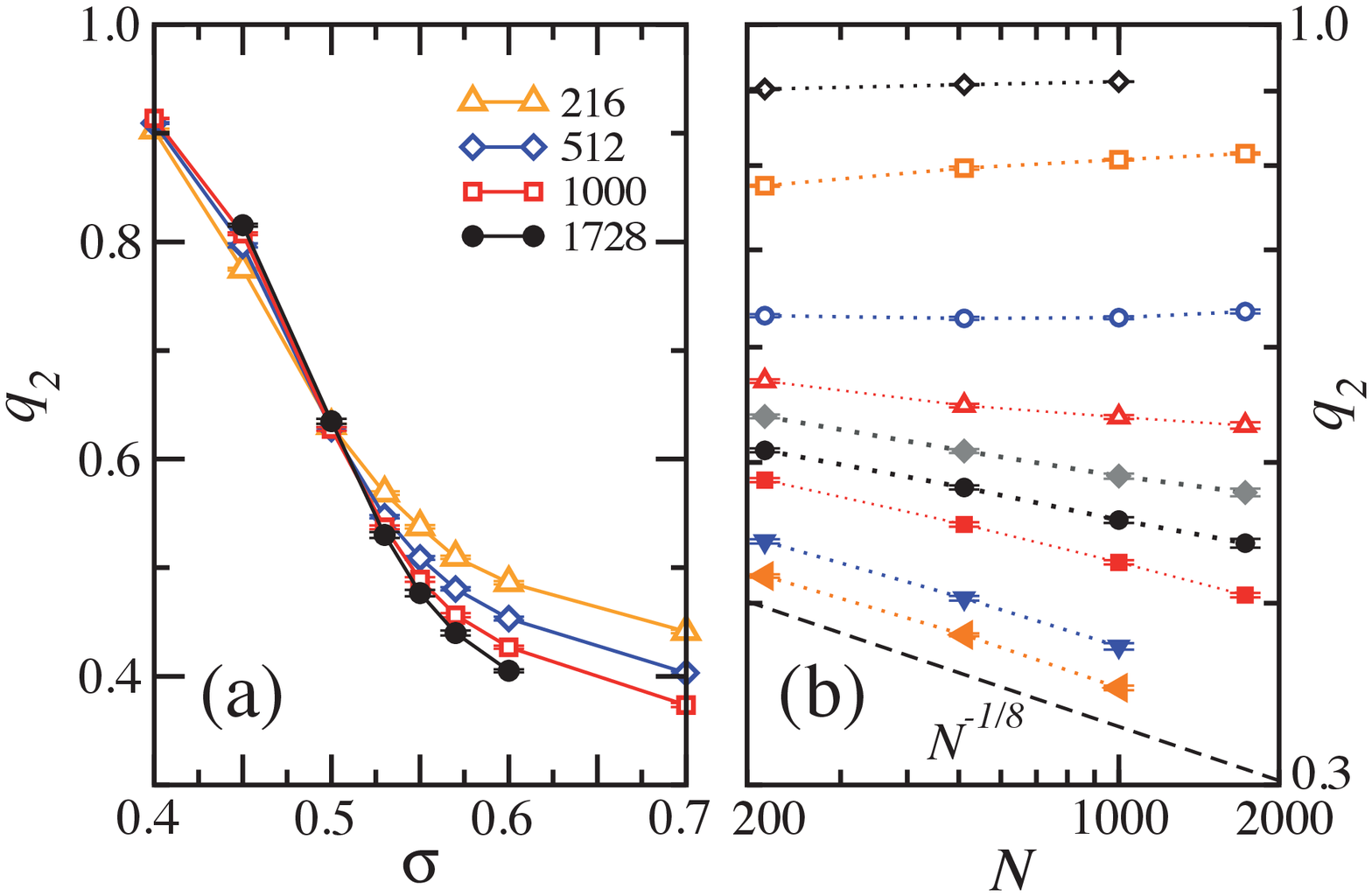}
\caption{ (Color online)
(a) Plots of the squared overlap parameter  $q_{2}$  vs $\sigma$ for the lowest available temperature, $T=0.55$.  $\smalltriangleup$,  
$\smalldiamond$, $\square$ and $\smallcircle$ stand for  $N=216, 512, 1000$ and $1728$ respectively. Solid lines are guides to the eye.
(b) Log-log plots of $q_2$  vs the number of dipoles $N$ for $T=0.55$. 
From top to bottom, $\smalldiamond$, $\smallsquare$, $\smallcircle$, $\smalltriangleup$, $\smallblackdiamond$, $\smallblackcircle$,
$\smallblacksquare$,  $\smallblacktriangledown$, and $\smallblacktriangleleft$ 
stand for $\sigma=0.4,~ 0.45,~ 0.5, ~0.53, ~0.55, ~0.57,~0.6, ~0.7$, and $\infty$ respectively.  
Dotted lines are guides to the eye. The dashed line corresponds  approximately to a $N^{-1/8}$ decay.}
\label{q2-SIG}
\end{center}
\end{figure}

The plots for $q_{2}$ are shown Fig.~\ref{q2-SIG}. Similarly as for $m_{2}$,
 $q_{2}$  does not vanish for $\sigma \le 0.53$, as it is expected for a FM phase.
For larger values of $\sigma$ we find instead a $1/N^{p}$ algebraic decay of $q_{2}$. 
Note that the slope of the decay is small. For example,  for $\sigma \ge 0.7$ we find $p \approx 1/8$, 
indicating that we are far from a PM phase (for which $p=1$ is expected).

We next examine how the cumulants $B_{m}$ and $B_{q}$ vary with $\sigma$ and $N$ at
low temperatures. For the FM phase, both quantities tend to $1$ in the thermodynamic limit while
for the SG phase $B_{m}$ should vanish as $N \to \infty$, and $B_{q}$ should tend to a non-zero value. 
Then, if there is a transition line separating the FM and the SG phases, we 
expect the related $B_{m}$ vs $\sigma$ curves to cross at the transition point $\sigma_{c}(T)$. As for the $B_{q}$ vs $\sigma$ curve, it should merge for $\sigma \ge \sigma_{c}$ and 
splay out only for $\sigma <  \sigma_{c}$.

\begin{figure}[!b]
\begin{center}
\includegraphics*[width=84mm]{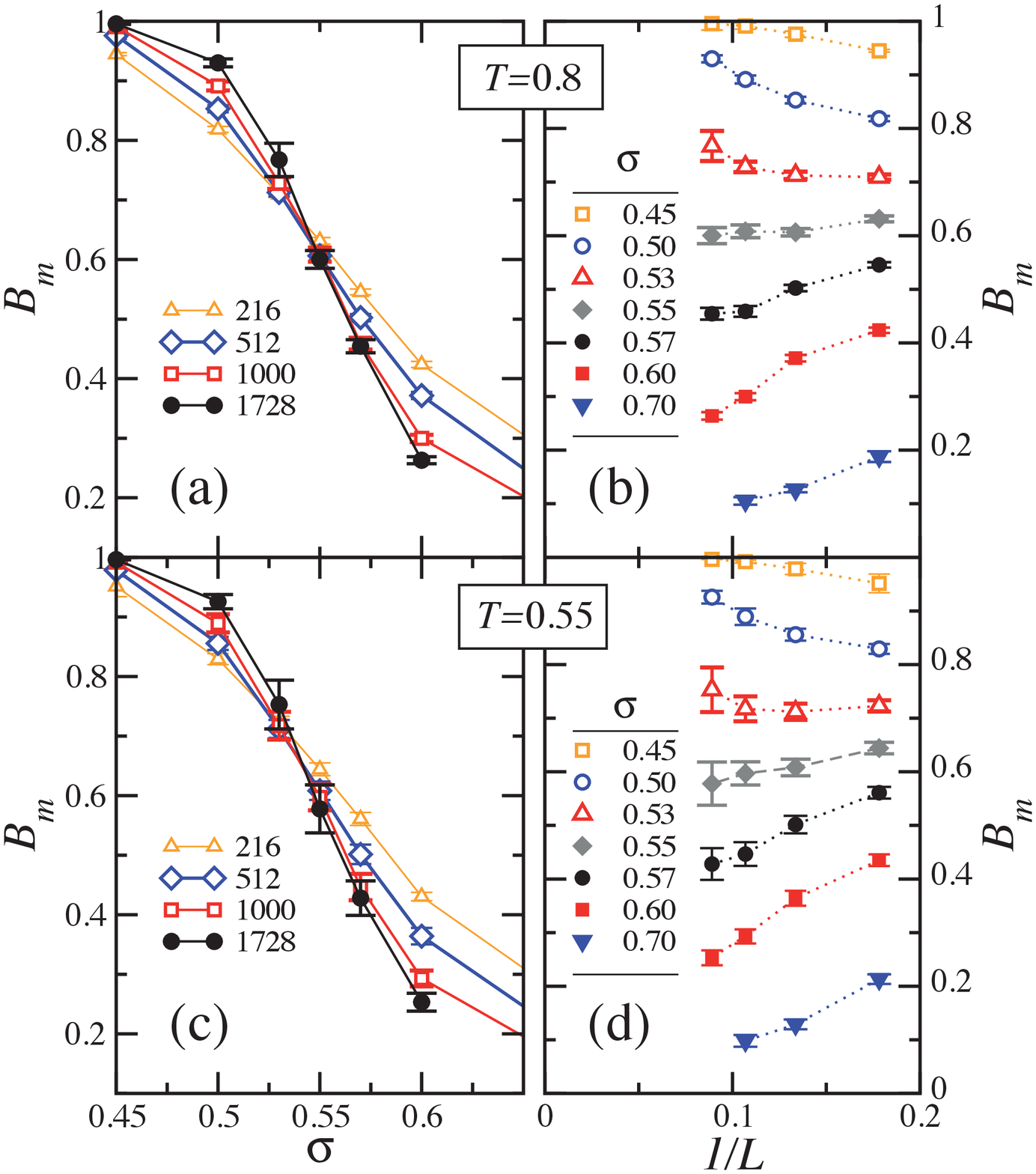}
\caption{
(Color online) (a) Plots of  $B_{m}$ vs $\sigma$ for $T=0.8$,
  and the values of  $N$ indicated in the panel.  Solid lines are guides to the eye.
  (b) Plots of  $B_{m}$ vs $1/L$  for $T=1$ for various values of $\sigma$.
  From top to bottom,  $\smallsquare$, $\smallcircle$, $\smalltriangleup$,
  $\smallblackdiamond$, $\smallblackcircle$, $\smallblacksquare$, and  $\smallblacktriangledown$ 
  stand for $\sigma=0.45,~ 0.5, ~0.53, ~0.55, ~0.57,~0.6$, and $0.7$ respectively.  
  Dotted lines are guides to the eye.   (c) Same as in (a) but for $T=0.55$.  (d) Same as in (b) but for $T=0.55$.} 
\label{bindersSIG}
\end{center}
\end{figure}

In Fig.~\ref{bindersSIG}(a) we show plots of $B_{m}$ vs $\sigma$  for $T=0.8$, a temperature that lies below the PM boundary.
Curves for different sizes do not cross at a precise point but rather tend to collapse 
in the intermediate region $0.55  \lesssim \sigma  \lesssim 0.57$ as $N$ increases. They
only splay out for $\sigma \lesssim 0.53$ and for $\sigma \gtrsim 0.6$. Plotting instead
$B_{m}$ vs $1/L$ for several values of $\sigma$, as shown in  Fig.~\ref{bindersSIG}(b), we see that
$B_{m}$ tends  to values that are neither $1$ nor $0$, which is a trait of quasi-long-range order,
only in this intermediate region. Similar plots are given for a lower temperature, $T=0.55$,  
in panels (c) and (d) of  the same figure. We obtain the same qualitative picture found for $T=0.8$, apart from the fact 
that finite size effects are larger within the intermediate region. However, $1/L$ extrapolations
of $B_{m}$ for $\sigma=0.55$ and  $0.57$ tend to non-vanishing values, which is consistent 
with marginal behavior.  We have performed averages over thousands of samples in order to improve  the statistics. However, the error 
bars of $B_{m}$ do not allow a precise determination 
of the FM boundary $\sigma_{c}(T)$. The points  along the FM boundary shown in Fig.~\ref{phases},
are  just rough estimates  obtained by taking the mean value of the crossing points of the pairs of curves
$B_{m}$ vs $\sigma$ for different sizes $(N_{1}, N_{2})=(8^{3},10^{3})$ and $(10^{3},12^{3})$. We find a 
boundary line which is nearly vertical with a positive slope suggesting a slight reentrance 
near $\sigma=0.55$. However, at least for the system sizes we have employed, plots of $m_{2}$ vs $T$ for $\sigma=0.55$ do 
not allow to discern any intermediate region with strong FM order separating the low temperature SG phase 
from the PM region (not shown). More extensive simulations for larger systems and for additional 
values of $\sigma$ within the interval $(0.53,0.6)$ would be needed to address this issue.
In summary, the results point to the existence of a narrow intermediate region with quasi-long-range order between the
FM boundary line and the SG phase, a phase which covers the low-temperature region for all $\sigma \ge 0.6$.
 For $\sigma=0.57$ and all temperatures below the PM boundary, we obtain 
a non vanishing $B_{m}$ and an algebraic decay of $m_{2}$ with $N$, indicating that that region of the $T$$-$$\sigma$ plane still stays in the quasi-long-range regime.
The area shaded with grey color in Fig.~\ref{phases} exhibits the extent of this intermediate phase.

  \begin{figure}[!b]
\begin{center}
\includegraphics*[width=84mm]{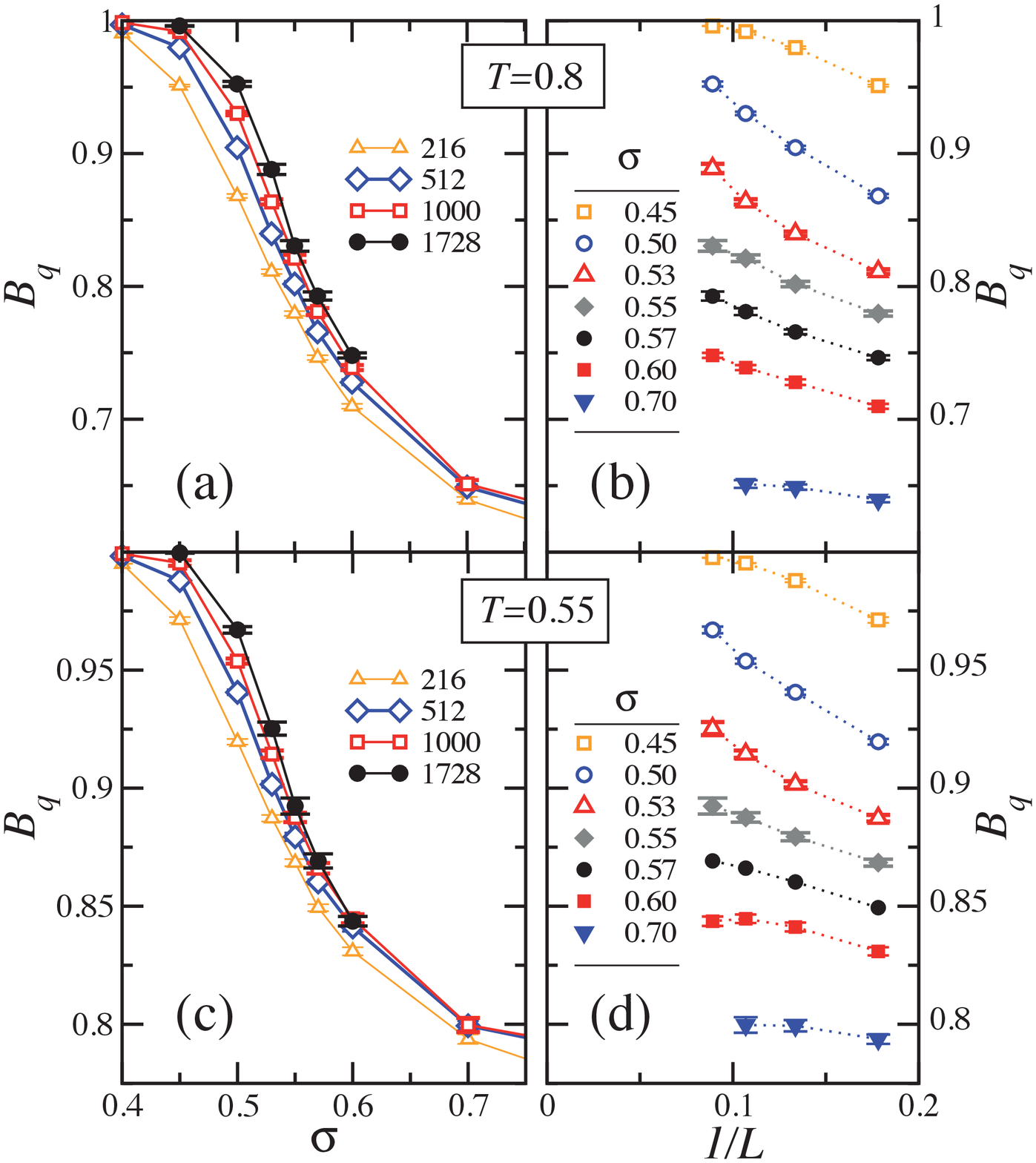}
\caption{
(Color online) (a) Plots of  $B_{q}$ vs $\sigma$ for $T=0.8$,
  and the values of  $N$ indicated in the panel.  Solid lines are guides to the eye.
  (b) Plots of  $B_{m}$ vs $1/L$  for $T=1$ for various values of $\sigma$.
  From top to bottom, $\smallsquare$, $\smallcircle$, $\smalltriangleup$, 
  $\smallblackdiamond$, $\smallblackcircle$, $\smallblacksquare$, and  $\smallblacktriangledown$ 
  stand for $\sigma=0.45,~ 0.5, ~0.53, ~0.55, ~0.57,~0.6$, and $0.7$ respectively.
  Dotted lines are guides to the eye.  (c) Same as in (a) but for $T=0.55$.  (d) Same as in (b) but for $T=0.55$.} 
\label{bindersQSIG}
\end{center}
\end{figure}

Additional information could be gathered from comparison of plots in
Fig.~\ref{bindersSIG} with their counterparts for $B_{q}$ vs $\sigma$ shown in Fig.~\ref{bindersQSIG}. 
Note that, in contrast to $B_{m}$, the curves of $B_{q}$ vs $\sigma$ do not splay out for $\sigma\ge 0.6$
but merge for large $N$. This is expected for the SG phase described in the previous section.
On the other hand, for $\sigma \le 0.53$ we find that both $B_{m}$ and $B_{q}$ tend to $1$ 
in the thermodynamic limit, indicating the existence of strong FM order. 
Finally, for $\sigma=0.55$ and $0.57$ (the only values we have simulated in the intermediate
region), $B_{q}$ increases with the size of the system. $1/L$ extrapolations of $B_{q}$ 
for $T=0.55$ point to values which are less than $1$, suggesting that the intermediate phase includes
quasi-long-range FM and SG order contemporaneously. Note however that the data for 
$T=0.8$ shown in Fig.~\ref{bindersQSIG}(b) do not exclude the possibility of having 
strong SG order in this intermediate region. Simulations for larger systems 
far beyond our present CPU-time resources would be needed in order to address this point.

\section{CONCLUSIONS }
\label{conclusion}

We have studied by Monte Carlo simulations the effect of  texturation on the collective behavior 
of disordered dense packings  of identical magnetic nanospheres that behave as Ising dipoles along local easy axes.
The local axes orientations follow a probability distribution parameterized by a single parameter $\sigma$.
This allows to vary the amount of orientational disorder ranging from the complete textured case ($\sigma=0$)
with all axes pointing along a common direction, to the non-textured 
one with the axes oriented at random ($\sigma=\infty$).

We have obtained the phase diagram on the temperature-$\sigma$ plane (see Fig. \ref{phases}), from studying the magnetization, the spin-glass overlap
parameter $q$, their fluctuations, as well as some other related observables, see \ref{meas}. The region $\sigma \le 0.53$ contains a
low-temperature ferromagnetic phase with strong order separated by a second order transition line from a paramagnetic high-temperature phase. For
large orientational disorder (namely, for $\sigma \ge 0.6$) the ferromagnetic order gives way to a
spin-glass phase for temperatures below  a nearly flat transition line $T_{sg}(\sigma)$ that extends up to $T_{sg}(\infty)=0.8$.
The spin-glass phase is similar to the one previously 
observed in systems of Ising dipoles
with strong structural disorder, at $\sigma =\infty$.  The Binder cumulants allow to estimate the position of the low-temperature boundary separating the
ferromagnetic and spin-glass phases. It is located near $\sigma=0.55$ and consistent with a small reentrance. Moreover, a narrow  intermediate 
region with quasi-long-range ferromagnetic order seems to lie between the ferromagnetic and the spin-glass phases.  

 Finally  we comment on the applicability of our results to actual experimental situations.
  As stated in the  introduction, the model corresponds to the limit $T_c/T_b \gg 1$ where $T_b$ is the blocking temperature of the dispersed system and $T_{c}$ a dipolar ordering temperature. This is for instance the situation of the maghemite NP ensembles with diameters  $d_p$ $6~{\rm nm} < d_p < 12~{\rm nm}$ studied in Ref. \cite{toro1}. In them, PM/SG freezing is observed for randomly distributed easy axes and a volume fraction $\phi$ ca. $0.67$ at a  ratio of temperatures
  $4 < T_c/T_b < 12$. Moreover the aging phenomenon used to characterize the SG state is observable only at temperatures  above $T_b$. We can thus conclude that the present model applies at a qualitative level to the latter experimental situations  whenever the SG region of the phase diagram is reached.
\\

\section*{Acknowledgements}

We thank the Centro de Supercomputaci\'on y Bioinform\'atica  at University of M\'alaga, 
Institute Carlos I at University of Granada and Cineca for their generous allocations of computer time in clusters Picasso, and Proteus.  We thank also access to the HPC resources of CINES under the allocation
2018-A0040906180 made by GENCI, CINES, France.
Work performed under grants FIS2017-84256-P (FEDER funds) from
the Spanish Ministry and the Agencia Espa{\~n}ola de
Investigaci{\'o}n (AEI), SOMM17/6105/UGR from Consejer\'{\i}a de Conocimiento, Investigaci\'on y Universidad, Junta de Andaluc\'{\i}a and European Regional Development Fund (ERDF), and ANR-CE08-007 from the ANR French Agency. J.J.A. also thanks the Italian ``Fondo FAI'' for financial support.

Each author also thanks the warm hospitality received during his stays in the other authors' institutes: ICMPE, the Pisa INFN section, and the  University of M\'alaga.


\begin{thebibliography}{99}
\bibitem{extraSG} For some values of $\sigma$, pairs of curves do not cross precisely at the same
point, but $1/L$ extrapolations of the crossing points allow to obtain proper values of $T_{sg}$.
\end{thebibliography}
\end{document}